\PassOptionsToPackage{dvipsnames,svgnames}{xcolor}
\documentclass[10pt,a4paper,onecolumn]{article}

\usepackage{marginnote}    
\usepackage{graphicx}      
\usepackage{xcolor}        
\usepackage{etoolbox}      
\usepackage[affil-it]{authblk} 
\usepackage{orcidlink}     
\usepackage{titlesec}      
\usepackage{calc}          
\usepackage{tikz}          
\usepackage{hyperref}      
\usepackage{caption}       
\usepackage{amssymb,amsmath} 
\usepackage{ifxetex,ifluatex} 
\usepackage{seqsplit}      
\usepackage{xstring}       
\usepackage{float}         

\let\origfigure\figure
\let\endorigfigure\endfigure
\renewenvironment{figure}[1][2] {
  \expandafter\origfigure\expandafter[H]
} {
  \endorigfigure
}

\makeatletter
 \let\@cite@ofmt\@firstofone
 \def\@biblabel#1{}
 \def\@cite#1#2{{#1\if@tempswa , #2\fi}}
\makeatother

\newlength{\cslhangindent}
\newlength{\csllabelwidth}
\newenvironment{CSLReferences}[2] 
 {\begin{list}{}{%
  \setlength{\itemindent}{0pt}
  \setlength{\leftmargin}{0pt}
  \setlength{\parsep}{0pt}
  \ifodd #1
   \setlength{\leftmargin}{\cslhangindent}
   \setlength{\itemindent}{-1\cslhangindent}
  \fi
  \setlength{\itemsep}{#2\baselineskip}}}
 {\end{list}}

\ifLuaTeX
  \usepackage[bidi=basic]{babel}
\else
  \usepackage[bidi=default]{babel}
\fi
\babelprovide[main,import]{american}

\DeclareUnicodeCharacter{2265}{\ensuremath{\geq}}   
\DeclareUnicodeCharacter{0308}{\"{}}

\let\textttOrig=\texttt
\def\texttt#1{\expandafter\textttOrig{\seqsplit{#1}}}
\renewcommand{\seqinsert}{\ifmmode
  \allowbreak
  \else\penalty6000\hspace{0pt plus 0.02em}\fi}

\makeatletter
\let\href@Orig=\href
\def\href@Urllike#1#2{\href@Orig{#1}{\begingroup
    \def\Url@String{#2}\Url@FormatString
    \endgroup}}
\def\href@Notdoi#1#2{\def\tempa{#1}\def\tempb{#2}%
  \ifx\tempa\tempb\relax\href@Urllike{#1}{#2}\else
  \href@Orig{#1}{#2}\fi}
\def\href#1#2{%
  \IfBeginWith{#1}{https://doi.org}%
  {\href@Urllike{#1}{#2}}{\href@Notdoi{#1}{#2}}}
\makeatother

\usepackage[top=3.5cm, bottom=3cm, right=1.5cm, left=1.0cm,
            headheight=2.2cm, reversemp, includemp, marginparwidth=4.5cm]{geometry}

\titleformat{\section}
  {\normalfont\sffamily\Large\bfseries}
  {}{0pt}{}
\titleformat{\subsection}
  {\normalfont\sffamily\large\bfseries}
  {}{0pt}{}
\titleformat{\subsubsection}
  {\normalfont\sffamily\bfseries}
  {}{0pt}{}
\titleformat*{\paragraph}
  {\sffamily\normalsize}

\usepackage{fancyhdr}
\makeatletter
\let\ps@plain\ps@fancy
\fancyheadoffset[L]{4.5cm}
\fancyfootoffset[L]{4.5cm}
\definecolor{linky}{rgb}{0.0, 0.5, 1.0}
\newcommand{\pkg}[1]{BH-VR}
\newcommand{\ExternalLink}{%
   \tikz[x=1.2ex, y=1.2ex, baseline=-0.05ex]{%
     \begin{scope}[x=1ex, y=1ex]
       \clip (-0.1,-0.1)
         --++ (-0, 1.2)
         --++ (0.6, 0)
         --++ (0, -0.6)
         --++ (0.6, 0)
         --++ (0, -1);
       \path[draw,
         line width = 0.5,
         rounded corners=0.5]
         (0,0) rectangle (1,1);
     \end{scope}
     \path[draw, line width = 0.5] (0.5, 0.5)
       -- (1, 1);
     \path[draw, line width = 0.5] (0.6, 1)
       -- (1, 1) -- (1, 0.6);
     }
   }

\patchcmd{\@maketitle}{center}{flushleft}{}{}
\patchcmd{\@maketitle}{center}{flushleft}{}{}
\patchcmd{\@maketitle}{\LARGE}{\LARGE\sffamily}{}{}
\def\maketitle{{%
  
  \AB@maketitle}}
\makeatletter
\renewcommand\AB@affilsepx{ \protect\Affilfont}
\renewcommand\AB@affilnote[1]{{\bfseries #1}\hspace{3pt}}
\renewcommand{\affil}[2][]%
   {\newaffiltrue\let\AB@blk@and\AB@pand
      \if\relax#1\relax\def\AB@note{\AB@thenote}\else\def\AB@note{#1}%
        \setcounter{Maxaffil}{0}\fi
        \begingroup
        \let\href=\href@Orig
        \let\texttt=\textttOrig
        \let\protect\@unexpandable@protect
        \def\thanks{\protect\thanks}\def\footnote{\protect\footnote}%
        \@temptokena=\expandafter{\AB@authors}%
        {\def\\{\protect\\\protect\Affilfont}\xdef\AB@temp{#2}}%
         \xdef\AB@authors{\the\@temptokena\AB@las\AB@au@str
         \protect\\[\affilsep]\protect\Affilfont\AB@temp}%
         \gdef\AB@las{}\gdef\AB@au@str{}%
        {\def\\{, \ignorespaces}\xdef\AB@temp{#2}}%
        \@temptokena=\expandafter{\AB@affillist}%
        \xdef\AB@affillist{\the\@temptokena \AB@affilsep
          \AB@affilnote{\AB@note}\protect\Affilfont\AB@temp}%
      \endgroup
       \let\AB@affilsep\AB@affilsepx
}
\makeatother

\renewcommand\Affilfont{\sffamily\small\mdseries}
\usepackage[T1]{fontenc}
\usepackage[utf8]{inputenc}

\IfFileExists{upquote.sty}{\usepackage{upquote}}{}
\IfFileExists{microtype.sty}{%
\usepackage{microtype}
\UseMicrotypeSet[protrusion]{basicmath} 
}{}

\usepackage{hyperref}
\hypersetup{unicode=true,
            pdfborder={0 0 0},
            breaklinks=true,
            pdflang={en-UK},
            colorlinks=true,
            linkcolor={Maroon},
            filecolor={Maroon},
            citecolor=[rgb]{0.0, 0.0, 1.0},
            urlcolor=linky,
            colorlinks,breaklinks=true,
            urlcolor=linky,
            linkcolor=linky}
\let\addcontentslineOrig=\addcontentsline
\def\addcontentsline#1#2#3{\bgroup
  \let\texttt=\textttOrig\addcontentslineOrig{#1}{#2}{#3}\egroup}
\let\markbothOrig\markboth
\def\markboth#1#2{\bgroup
  \let\texttt=\textttOrig\markbothOrig{#1}{#2}\egroup}
\let\markrightOrig\markright
\def\markright#1{\bgroup
  \let\texttt=\textttOrig\markrightOrig{#1}\egroup}
  
\IfFileExists{parskip.sty}{%
\usepackage{parskip}
}{
\setlength{\parindent}{0pt}
\setlength{\parskip}{6pt plus 2pt minus 1pt}
}
\ifx\paragraph\undefined\else
\let\oldparagraph\paragraph
\renewcommand{\paragraph}[1]{\oldparagraph{#1}\mbox{}}
\fi
\ifx\subparagraph\undefined\else
\let\oldsubparagraph\subparagraph
\renewcommand{\subparagraph}[1]{\oldsubparagraph{#1}\mbox{}}
\fi

\title{Investigating encounters between black holes in gravitational NBODY simulations of globular clusters using \pkg{}}

\author[1,2]{Gaku Kawashima%
    \,\orcidlink{0009-0006-1396-8397}\,%
    }
\author[1,3,4]{Lucy O. McNeill%
    \,\orcidlink{0000-0003-1752-4882}\,%
    }
\author[1]{Herman Shiu-Hang Lee %
    \,\orcidlink{}\,%
    }

\affil[1]{Department of Astronomy, Kyoto University, Kyoto 606-8502, Japan\\
}%
\affil[2]{Center for Astrophysics | Harvard \& Smithsonian, 60 Garden Street, Cambridge, MA 02138, USA
}%
\affil[3]{Hakubi Center, Kyoto University
}%
\affil[4]{Center for Interdisciplinary Theoretical and Mathematical Sciences (iTHEMS), RIKEN, Saitama 351-0198, Japan
}%

\hypersetup{pdftitle={Code paper name},
            pdfauthor={Author1 Name, Author2 Name}}

\begin{document}
\date{}
\maketitle

\marginpar{
  \begin{flushleft}
  \sffamily\small



  {\bfseries Software}
  \begin{itemize}
    \setlength\itemsep{0em}
    \item \href{https://github.com/gkawa08/BH-VR}{\color{linky}{Repository}} \ExternalLink
  \end{itemize}

  \vspace{2mm}
  \par\noindent\hrulefill\par
  \vspace{2mm}


  {\bfseries Published:} unpublished

  \vspace{2mm}
  {\bfseries License}\\
  Authors of papers retain copyright and release the work under a Creative Commons Attribution 4.0 International License (\href{https://creativecommons.org/licenses/by/4.0/}{\color{linky}{CC BY 4.0}}). 
  \end{flushleft}
}

\vspace{-0.8cm}

\section{Summary}
\label{summary}

\pkg{} is an open-source Virtual Reality (VR) application for visualising the gravitational dynamics of black holes (BH) in globular clusters. While immersed in a gravitational NBODY6++ simulation, users can investigate successive ``strong'' gravitational interactions between binary black holes (BBH) in the globular cluster core. This includes the gravitational wave driven ``MERGER'' of BBHs, and also the successive gravitational three-body ``EXCHANGE'' interactions between BBH + single BH interlopers which act to shrink the binary separation preceding a MERGER. 
\pkg{} streamlines data analysis of NBODY output to directly observe the formation of so--called ``hierarchical'' BBH mergers which can result from successive three body EXCHANGE and MERGER events which involve black holes forbidden by stellar evolution theory e.g. GW231123.

\section{State of the field}
\subsection{Astrophysics context} \label{context}
Since 2015, the gravitational wave driven mergers of binary black holes (BBH) at extra-Galactic distances ($\sim$Gpc) are being detected by the LIGO Virgo KAGRA (LVK) network (\cite{Abbott2016}). At present, over 300 BBH mergers have been reported (\cite{LVK2026}). A minority ($\sim 1 \%$) of events challenge the conventional understanding of stellar evolution because they contain exceptionally massive and/or highly spinning black holes which cannot be produced by massive stars in ``isolated'' binaries. The most extreme example is the BBH merger GW231123  (\cite{Abac2025}) which had a BBH merger product mass of 190-265~$M_\odot$. Theoretically, globular clusters are popular candidate sites for producing such massive and highly spinning black holes, since their high densities facilitate the possibility for several generations (second, third etc.) of ``hierarchical'' BBH mergers. This scenario allows the formation of black holes in the ``mass-gap'' due to pair-instability supernova explosions, which forbids the formation of black holes from single stars between 50--130~$M_\odot$ (e.g., \cite{Heger2002}). 
\\

Unlike in isolated environments, black holes in globular clusters gravitationally interact with each other. For massive ($\gtrsim 10^5$~$M_\odot$) clusters with hundreds of BHs, ``strong'' gravitational interactions occur between BBH+BH in the cluster core relatively frequently, once every 10--100~Myr due to their large interaction interaction cross section. In BBH+BH strong interactions, after temporarily existing as a bound three-body system (lower center panel of Figure~\ref{fig:flow}), one of the two binary components (pink) can be swapped out for the ``interloper'' (red). Successive three body encounters between BBH and BH act to harden BBHs, which impart dynamical recoil kicks on both the post-interaction BBH and BH. 
\subsection{Computational context} \label{statement-of-need}
Gravitational NBODY simulations have been used to model stellar dynamics in globular clusters for 50 over years (\cite{Aarseth1971,Aarseth2003}). Simulations are computationally expensive, requiring months on a supercomputer even for clusters with $\sim100,000$ stars. Recently, the formation of GW231123--like BBH mergers has been borne out by gravitational NBODY simulations of 100,000--millions of stars thanks to developments in the NBODY6++GPU code (\cite{Nitadori2012,Wang2020}), such as the 7th generation BBH merger resulting in a 100~$M_\odot$ black hole in \cite{Anagnostou2022}.
\section{Statement of need}\label{purpose}
\pkg{} streamlines the identification of the BBH+BH strong encounters and BBH mergers that may eventually lead to a GW231123--like BBH merger event from an NBODY globular cluster simulation. By handling the specialized data manipulation required to extract this information, \pkg{} removes this bottleneck to immediately visualize and investigate the dozens of BBH+BH exchange and BBH merger events.
\section{Software Design}\label{software-design}
\begin{figure}
    \centering
\includegraphics[width=0.8\linewidth]{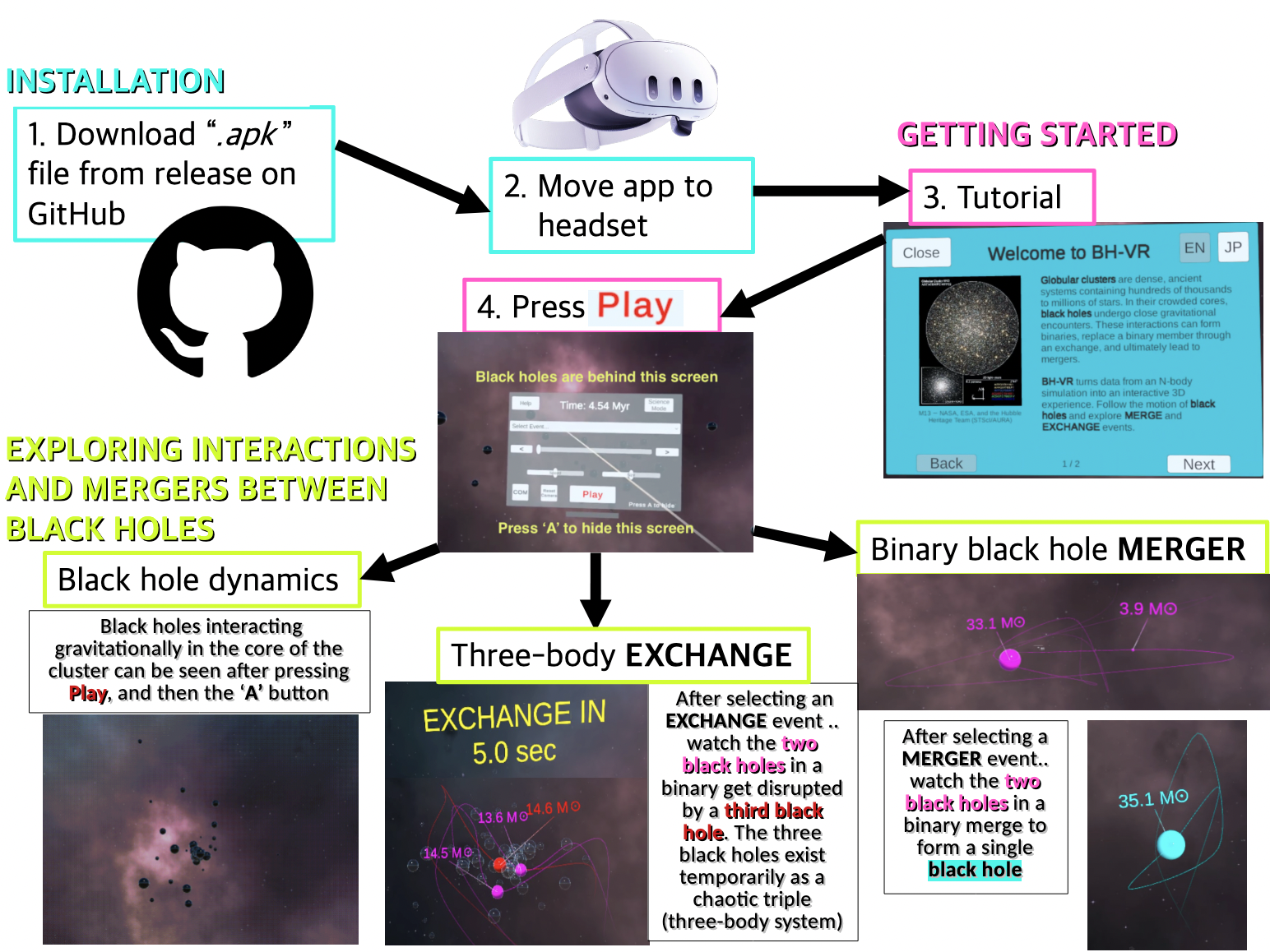}
    \caption{Overview of the  installation (cyan), preparation (magenta) and scientific use (green) of \pkg{}. Details for respective steps are on GitHub. Once installed (cyan and magenta), it is possible to view all black holes in the cluster interacting gravitationally in the cluster core (green, bottom left). It is also possible to select any binary+single (or 2+1) EXCHANGE interactions (bottom center). When applicable, users can watch any gravitational wave driven BBH merger event that happened in the simulation (bottom right).}
    \label{fig:flow}
\end{figure}

The most basic usage case for running \pkg{} on a Unity headset is shown in Figure~\ref{fig:flow}.
First, users install the GitHub release, which is contained in a single \textit{.apk} file (1.). Once in the headset (2.), users can launch the app, where a blue Tutorial page will appear (3.). Users can click through by pressing the ``Next'' button (with the back right trigger of the control). Finally, the home menu is shown (4., central panel). By pressing ``Play'' (and then hiding the menu with the ``A'' button) users can view the gravitational dynamics of the black holes in the cluster over 14~Gyr. 
\\

Returning to the menu with the ``A'' button, users can change the speed of the simulation (bottom left slider), or manually move the time slider (upper slider). In addition, users can change the relative size of the black holes (which each have a radius $R_\mathrm{BH}\propto m_\mathrm{BH}$ to be physically accurate).
\\

The drop down menu at the top contains a chronological list of all three body ``EXCHANGE'' interactions (lower middle part of flow chart), where a BBH (magenta spheres) is disrupted by an interloper (red sphere). As the interloper approaches the binary, the three BH are gravitationally bound and the system exists temporarily as an unstable triple system. Users can watch and move around this chaotic three body interaction. Eventually, one of the three members is ejected, leaving a BBH binary with a new semi-major axis and possibly different mass components. \pkg{} also shows any BBH ``MERGER'' in the simulation. These typically occur after a few ``EXCHANGE'' interactions. Users can view the BBH (magenta spheres) which has a relatively short orbital period of $\sim$minutes. Once the BBH merges, the remnant black hole is in cyan.
\\

In order to investigate the precise sequence of events leading up to e.g. the formation of hierarchical mergers (such as e.g.  GW231123), users can click on ``Science Mode'' in the top right of the menu. When users select an ``EXCHANGE'' or ``MERGER'' while in Science Mode, all relevant physical properties are shown. Science Mode also includes a ``Follow'' mode where users can select a specific black hole ID and move to its reference frame. Here, users assume the position of the black hole in space as it interacts gravitationally with other BH and BBH.

\section{Research Impact Statement}\label{research-impact-statement}
It has been suggested that the majority of BBH merger events may come from globular clusters (\cite{Rodriguez2015}). However, it is computationally difficult to survey the optimum properties (number of stars, metallicity, initial binary fraction etc.) that produce the largest BBH merger rate with direct NBODY simulations. There are recent efforts to approximate the evolution of black holes in globular clusters with simplified models (e.g., \cite{Kritos2024}). However, these emulators predict expectation values of e.g. the merger rate, and do not capture the unlikely, but possible, conflux of hardening encounters and dynamical recoil kicks that lead to a BBH merger inside the cluster. As an example, \cite{Anagnostou2022} made seven realisations of the same cluster i.e. with identical initial conditions. In one particular simulation, stochastic EXCHANGE and MERGER events lead to a 7th generation BBH merger within 14~Gyr whose final product is $\sim 100$~$M_\odot$.
With Science Mode, users can interactively piece together these lucky sequences of events which may represent the formation of BBH detected by the LVK network. This stochasticity can be understood by the following two properties listed in Science Mode:
\begin{enumerate}
\item \textbf{Binary separation before and after exchange: } NBODY emulators assume that EXCHANGE encounters reduce the binary orbital separation $a_\mathrm{before}$ by a constant factor based on the three mass components (e.g. $a_\mathrm{after}/a_\mathrm{before} \approx 0.76$ in the equal mass limit, \cite{Quinlan1996}). However, following a chaotic binary+single encounter, the new binary semi-major axis $a_\mathrm{after}$ is drawn from a \textit{distribution} (\cite{Stone2019}), which in the equal mass limit follows $P(a_\mathrm{after}/a_\mathrm{before})\propto (a_\mathrm{after}/a_\mathrm{before})^{3/2}$. In other words, much more extreme hardening $a_\mathrm{after}/a_\mathrm{before}\sim0.1$ is possible (though unlikely). With \pkg{} users can investigate the series of stochastic $a_\mathrm{after}/a_\mathrm{before}$ that precede BBH mergers.
\item \textbf{Masses of the binary before and after an EXCHANGE:} In NBODY emulators, it is assumed that for any binary+single EXCHANGE encounter, the post-interaction binary contains the two most massive components. However, due to the stochastic nature of three body encounters, there is a non-zero probability that the least massive component actually remains in the binary (\cite{Ginat2021}). This is the second feature of 2+1 BBH encounters that is not being captured by emulators. Observationally, there is a diverse mass ratio $q$ in the BBH merger population. But cluster emulators assume that the post-interaction BBH contains the most massive BHs, favoring more equal $q$ for hierarchical mergers. Emulator results may therefore (incorrectly) suggest that low $q$ BBH merger events are not associated with dynamical formation, instead favoring the isolated channel. This is particularly relevant for GW recoil kicks during BBH mergers which may result in the product BH being ejected from the cluster. Products from large $q>>1$ BBH are more likely to be retained (\cite{Campanelli2007}). With \pkg{}, it is possible to track the precise masses involved in EXCHANGES and MERGERS in the cluster that lead to a very unequal mass $q\sim 0.1$ BBH merger like GW190814. 
\end{enumerate}
\pkg{} also shows 3. \textbf{The location of a BBH MERGER or EXCHANGE event relative to the cluster core}: if these events happen far from the core, it is easier to eject a BBH or BH from a dynamical recoil kick, or a GR recoil kick. 
In summary, \pkg{} will be utilized to understand the stochastic nature of three body black hole encounters that lead to some of the most puzzling and exotic BBH mergers.
\section{AI usage disclosure}\label{ai-usage-disclosure}

Gemini and ChatGPT were used for the development of this app. The most significant AI contributions were the assistance of writing the Csharp and python codes. This manuscript was prepared without the use of AI.

\section{Acknowledgements}\label{acknowledgements}
This research was funded by the Hakubi Centre of Advanced research at Kyoto University. We thank Michele Trenti for kindly allowing us to use the NBODY6 simulations performed on HPC clusters at the University
of Melbourne (Spartan) and at the Swinburne University (OzSTAR). Swinburne University's OzStar supercomputing system was used to analyze the simulations. LM is grateful to Ilya Mandel for many years of fruitful discussions which inspired this software. We thank Raphael Hirschi and Max Isi for valuable feedback. 
\newpage
\phantomsection
\section{References}
\label{refs}

\end{document}